# Timing performance of 30-nm-wide superconducting nanowire avalanche photodetectors

F. Najafi[1†], F. Marsili[1†], E. Dauler[2], R. J. Molnar[2], K. K. Berggren[1‡]

[1] *Department of Electrical Engineering and Computer Science, Massachusetts Institute of Technology, 77 Massachusetts Avenue, Cambridge, Massachusetts 02139, USA*

[2] *Lincoln Laboratory, Massachusetts Institute of Technology, 244 Wood St., Lexington, Massachusetts 02420, USA*

We investigated the timing jitter of superconducting nanowire avalanche photodetectors (SNAPs, also referred to as cascade-switching superconducting single-photon detectors) based on 30-nm-wide nanowires. At bias currents ($I_B$) near the switching current, SNAPs showed sub-35-ps FWHM Gaussian jitter similar to standard 100-nm-wide superconducting nanowire single-photon detectors. At lower values of $I_B$, the instrument response function (IRF) of the detectors became wider, more asymmetric, and shifted to longer time delays. We could reproduce the experimentally observed IRF time-shift in simulations based on an electrothermal model, and explain the effect with a simple physical picture.

Superconducting nanowire avalanche photodetectors (SNAPs, also referred to as cascade-switching superconducting single-photon detectors) [1] are based on a parallel-nanowire architecture (Figure 1a) that allows single-photon counting with higher signal-to-noise ratio (up to a factor of ~ 4 higher [2]) than superconducting nanowire single-photon detectors (SNSPDs) [3] with the same nanowire width. Figure 1b shows the equivalent electrical circuit of a SNAP with 4 parallel sections (or 4-SNAP). All of the sections have nominally the same kinetic inductance ($L_0$) and are connected in series with an inductor ($L_S$) and in parallel with a readout resistor ($R_{load}$). If the bias current ($I_B$) of a $N$-SNAP is higher than the avalanche threshold current ($I_{AV}$) of the device, when one section switches to the normal state after absorbing a photon (initiating section), it diverts its current to the remaining $N$-1 sections (secondary sections), driving them normal (we call this process an avalanche). Therefore, a current ~ $N$ times higher than the current through an individual section is diverted to the read-out [2].

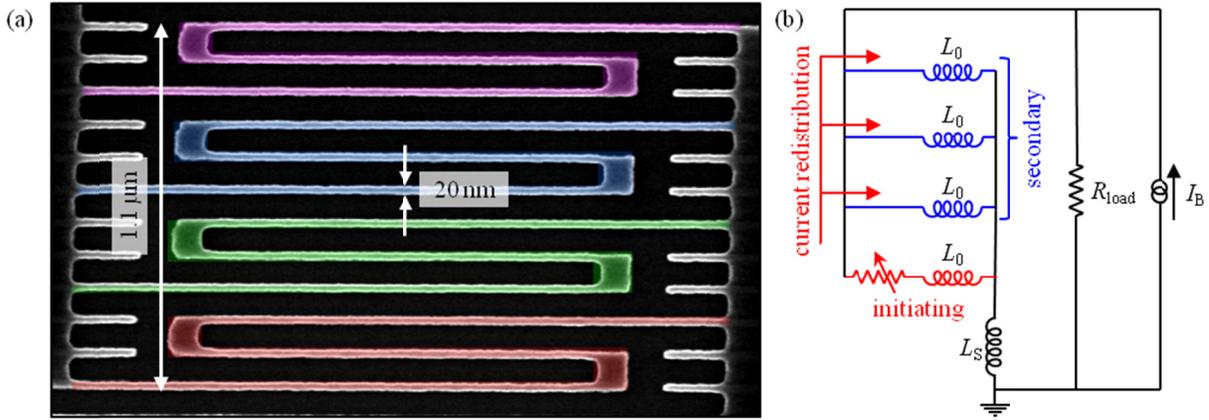

**Figure 1. a.** Colorized SEM image of a 4-SNAP resist (hydrogen silsesquioxane) mask on NbN with each section colored differently. **b.** Equivalent electrical circuit of a 4-SNAP. The arrows pointing at the secondary sections represent the current redistributed from the initiating section to the secondary sections after the initiating section switches to the normal state.

The physical origin of the photodetection delay and timing jitter of detectors based on superconducting nanowires remains unclear over 10 years after the introduction of these detectors. Zhang et al. [4] studied the photodetection delay of 130-nm-wide nanowires as a function of power and hypothesized that the observed 70-ps decrease of photodetection delay between the single-photon and multi-photon regimes might be due to reduced gap suppression time in the multi-photon regime. O'Connor et al. [5] studied the local dependence of the photodetection delay and timing jitter (190 – 205 ps) along 100-nm-wide nanowires and concluded that narrower nanowire sections have lower delay and jitter. However, significantly lower jitter values (~30 ps [6] to ~60 ps [7]) have been repeatedly reported for 100-nm-wide nanowires. Another study of jitter as a function of wavelength [8] found no dependence in the range 1 – 2 µm. Along with the dependence on nanowire width, incident optical power and photon energy, bias-current-dependence may provide decisive insight into the physical origin of photodetection delay and jitter. However, jitter measurements as a function of $I_B$, which have not been reported so far, have been hampered by decreasing signal-to-noise ratio (SNR, which makes the jitter induced by the electrical noise of the set up dominant over the jitter of the device) and exponentially decreasing detection efficiency (which makes the acquisition time of the instrument response function significantly longer [9]) with decreasing $I_B$. We recently found a way to overcome these obstacles: we employed SNAPs to read out 20- and 30-nm-wide nanowires [2]. The detection efficiency at 1550 nm wavelength was 17-20% and showed only a weak bias-current-dependence (<5% relative variation) in the bias range $I_{AV} < I_B < I_{SW}$, where $I_{SW}$ is the SNAP switching current [2]. Taking advantage of the possibility of efficiently detecting single photons over the entire SNAP bias range with high SNR ( > 3, as defined in Ref. [2]), we studied the timing performance of 30-nm-wide 2-, 3- and 4-SNAPs as a function of the bias current. Our results suggest that the gap suppression time, which would be expected to be strongly dependent on the bias current, has little if any effect on the most-likely photodetection delay when the detectors are operating in single-photon regime.

We measured the instrument response function (IRF) of 10 devices with active areas ranging from 0.8 to 2.1 μm² (see Ref. [2] for details on the fabrication process). Our main finding is that, although at bias currents near $I_{SW}$ the IRF of SNAPs had a Gaussian shape with sub-35-ps full width at half maximum (FWHM), at lower values of $I_B$ the IRF became wider, more asymmetric, and shifted to longer time delays. We could simulate the experimentally observed IRF time-shift (but not the observed asymmetry) by using an electrothermal model [10].

To illuminate the detectors, we used a mode-locked, sub-ps-pulse-width laser emitting at ~1550 nm wavelength with 77 MHz repetition rate. The laser output was split into two single-mode optical fibers that we coupled to the detector under test and to a low-jitter fast photodiode (pulse rise time < 35 ps). The signals from the SNAP and from the fast photodiode were sent to a 6-GHz-bandwidth, 40-GSample/s oscilloscope, which we used to measure the IRF. We verified that the SNAPs were operating in the single-photon regime by setting the power level of the incident light within a range in which the detector photoresponse counts increased linearly with incident power (as in Ref. [2]).

Figure 2a schematically represents the moments of the photodetection process most relevant to our discussion: (1) $t_0$: a sub-ps laser pulse is emitted; (2) $t_{FPD}$: the rising edge of the photoresponse pulse of the fast photodiode crosses the oscilloscope trigger level set to 50% of the average pulse peak value; (3) $t_{HSN}$: a photon is absorbed in the nanowire and it starts a resistive state formation process (hotspot nucleation, HSN); (4) $t_\xi$: the first resistive slab of length $\xi$ (the coherence length of NbN [10]) is formed across the width of the initiating section; (5) $t_{SNAP}$: the rising edge of the SNAP photoresponse pulse crosses the oscilloscope trigger level set to 50% of the average pulse peak value (which depends on $I_B$); and (6) $t_{95\%}$: the SNAP photoresponse pulse crosses the oscilloscope trigger level set to 95% of the average pulse peak value.

We defined the detector IRF as the histogram of the time delay $t_D$ measured on the oscilloscope between the rising edges of the fast photodiode pulse ($t_{FPD}$) and of the SNAP pulse ($t_{SNAP}$), i.e. $t_D = t_{SNAP} - t_{FPD}$. The IRF histograms were calculated by using $\sim 6 \cdot 10^4$ time delay samples. The absolute value of $t_D$ was set by the propagation times of the signals (laser pulse, fast-photodiode pulse and SNAP pulse) through the optical and electrical paths of our set up, as illustrated by arrows in Figure 2a. These delays were irrelevant to the problem. Therefore, for convenience we added an offset [11] so that $t_D = 0$ s at the maximum of the IRF when the device under test was biased at $I_B = I_{SW}$.

Figure 2b shows the IRF of a 2-SNAP (normalized by its maximum value) at different bias currents. We observed two current-dependent effects in the IRF: (1) as $I_B$ was increased, the time delay corresponding to the maximum of the IRF (we called this time delay "maximum-likelihood delay",

MLD) shifted towards shorter time delays; (2) as $I_B$ was decreased, the IRF progressively transitioned from a Gaussian shape (when the detector was biased close to $I_{SW}$) to a broader and asymmetric shape, exhibiting a decaying tail which extended for several hundreds of picoseconds beyond the MLD.

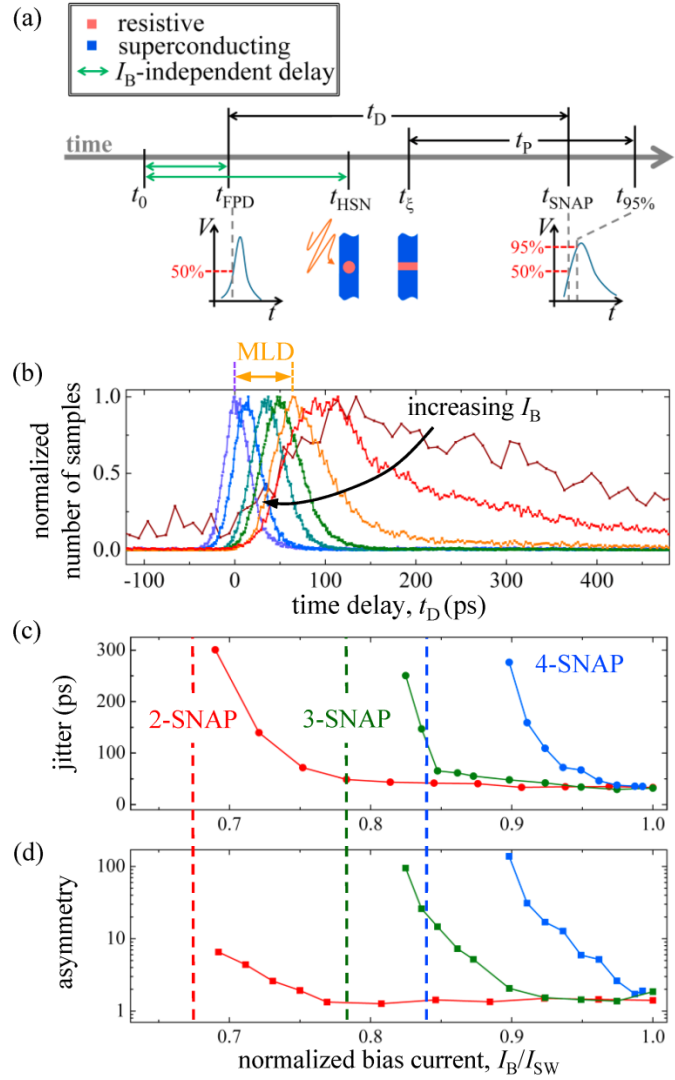

**Figure 2. a.** Schematic representation of instances during the photodetection process. A photon from an optical pulse emitted at $t_0$ is absorbed in the initiating section ($t_{HSN}$), generating a resistive slab along the width of the nanowire ($t_\xi$). After the avalanche, the SNAP bias current is diverted into the load and an output voltage pulse forms across the load resistor. The arrival of this pulse can be detected once the rising edge of the SNAP pulse crosses the trigger level of the oscilloscope ($t_{SNAP}$). We measured the time delay between $t_{SNAP}$ and a reference $t_{FPD}$, the instant at which the rising edge of the photodetection signal from a fast photodiode crossed the trigger level of the oscilloscope. The voltage ($V$) vs time ($t$) curves

represent the oscilloscope traces of the fast photodiode (left hand side) and SNAP (right hand side) pulses. The dashed lines represent the 50% and 95% thresholds. **b.** IRF (normalized by the maximum of each trace) of a 30-nm-wide 2-SNAP at bias currents: $I_B / I_{SW}$ = 1; 0.93; 0.85; 0.78; 0.73; 0.69 and 0.64. The curved arrow indicates the direction of increasing $I_B$. The double-pointed arrow indicates the maximum-likelihood delay (MLD) at $I_B / I_{SW}$ = 0.73. The MLD of the IRF was set to 0 seconds at $I_B / I_{SW}$ = 1. **c.** Jitter of a 2-, 3- and 4-SNAP based on 30-nm-wide nanowires as a function of the normalized bias current ($I_B / I_{SW}$). The switching currents of the 2-, 3-, and 4-SNAP were 13.2 µA, 17.9 µA, and 27.8 µA respectively. The vertical dashed lines indicate the avalanche currents of the SNAPs [2]. The data for the jitter of 3- and 4-SNAPs biased below $I_{AV}$ are not shown (see Ref. 12) as the devices were not operating as single-photon detectors (they were instead operating in arm-trigger regime as described in Ref. 2 ). **d.** IRF asymmetry vs. $I_B / I_{SW}$ for the same devices shown in panel c.

Figure 2c shows the jitter of 2-, 3- and 4-SNAPs, defined as the FWHM of the IRF, as a function of $I_B/I_{SW}$. The jitter of SNAPs showed a weak dependence on the bias current for $I_B$ close to $I_{SW}$ (e.g. for a 2-SNAP the jitter increased from 35 ps at $I_B = 0.97I_{SW}$ to 41 ps at $I_B = 0.88I_{SW}$), but rapidly increased as $I_B$ approached $I_{AV}$ (by ~ 100 ps for a decrease in $I_B$ of $0.1I_{SW}$). $I_{AV}$ was determined from detection efficiency measurements, as reported in Ref. [2]. We note that for $I_B$ approaching $I_{SW}$, SNAPs showed the same jitter as standard SNSPDs [6] (~33 ps), in contrast to previous reports of larger timing jitter of SNAPs [13,14].

Figure 2d shows the IRF asymmetry, defined as the ratio between the length of the IRF tails (experimentally defined as the time between 90% and 10% of the IRF maximum) after and before the MLD. Like the jitter, the asymmetry of SNAPs showed a weak dependence on the bias current at high $I_B$, but rapidly increased as $I_B$ approached $I_{AV}$.

The shift of the MLD to shorter delays with increasing $I_B$ can be explained by considering the dependence of the electrothermal dynamics of the device on the bias current. Using the electrothermal model described in Ref. [10], we simulated the time evolution of the current diverted from the SNAP to the read out ($I_{out}$) after a HSN event occurred in the initiating section. Our model did not describe the formation and expansion of the photon-induced hotspot [15,16], so in our simulations the absorption of a photon resulted in the immediate formation of a resistive $\xi$-long slab ($\xi$-slab), i.e. $t_\xi = t_{HSN}$.

We repeated the simulation at different values of $I_B$. Figure 3a shows the simulated current pulses from a 2-SNAP. We defined the detector peak delay $t_P$ as $t_P = t_{95\%} - t_\xi$, and set $t_\xi$ to 0 s in our simulations (see Ref. [12] for details on the choice of $t_{95\%}$ as reference). The observed increase of $t_P$ with decreasing $I_B$ can be easily understood. After the avalanche, the resistance $R(t)$ of the SNAP grows with time [10] at a rate that monotonically increases with the dissipated power proportional to $R(t) \cdot I_B^2$ (Joule heating). At lower bias currents, the dissipated power is smaller, resulting in a slower increase of $R(t)$. Hence, it takes longer for the diverted current $I_{out}$ to reach its peak value. Figure 3b shows $t_P$ (simulation) and the MLD (experiment) of a 2-SNAP as a function of $I_B$. As $t_P$ and the MLD show a similar dependence on $I_B$, and recalling that our choice of origin of the MLD was arbitrary, we conclude that the MLD differs at most by a current-independent offset from $t_P$.

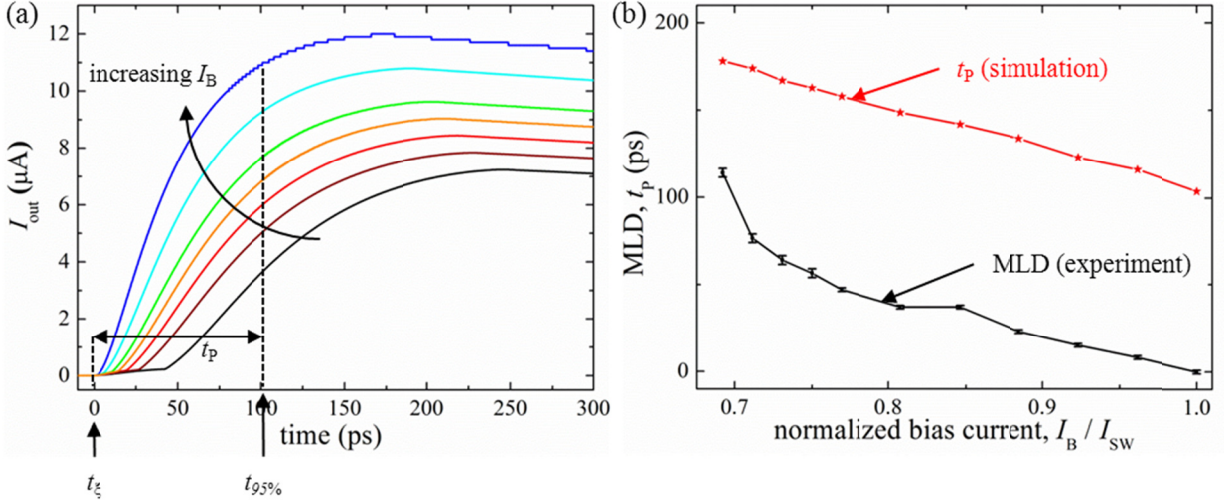

**Figure 3. a.** Simulated time evolution of the current diverted to the read out resistor ($R_{load} = 50\ \Omega$) by a 2-SNAP after a resistive $\xi$-slab is formed in the initiating section (at time $t_\xi = 0$ s) for $I_B / I_{SW} = 0.96, 0.87, 0.81, 0.77, 0.73, 0.70, 0.66$. The kinetic inductance of each section of the 2-SNAP was $L_0 = 13$ nH and the series inductor was $L_S = 130$ nH, corresponding

to the device of Figure 2b. Arrows indicate the time at which the resistive $\xi$-slab is formed ($t_\xi$); the time at which $I_{out}$ reaches 95% of its maximum ($t_{95\%}$) for $I_B / I_{SW} = 0.96$; the detector peak time for $I_B / I_{SW} = 0.96$ ($t_P$); and the direction of increasing $I_B$. **b.** Experimental MLD vs $I_B$ (squares) and simulated $t_P$ vs $I_B$ (stars) for the 2-SNAP of Figure 2b. The error on the MLD values was assumed to be twice the bin size of the IRF histograms. The value of the MLD for the highest $I_B$ was set to 0 s.

The absolute values of the MLD and of $t_P$ were defined with respect to different moments in time ($t_{FPD}$ for the MLD and $t_\xi$ for $t_P$) and by using different thresholds on the SNAP photoresponse pulse (50% of the average pulse peak value for the MLD and 95% of the average pulse peak value for $t_P$, see Ref. [12]). However, the current dependencies of the values of the MLD and $t_P$ were similar across the entire bias-current range (~ 30% of $I_{SW}$) of single-photon operation. From the comparison between the experimentally measured MLD values and the calculated $t_P$ values, we conclude that, when neglecting the effect of jitter, the MLD dependence on current can be entirely accounted for by the bias dependence of the peak delay $t_P$. Therefore, time difference $t_\xi - t_{FPD}$ does not significantly contribute to the bias dependence of the MLD. Since the time difference $t_{HSN} - t_{FPD}$ does not depend on $I_B$ by definition [17], we can conclude that $t_\xi - t_{HSN}$ does not appreciably affect the bias dependence of the MLD either. Therefore, the gap suppression time [4], which is known to be current-dependent [18], has negligible influence on the resistive-slab creation time, i.e. the time difference $t_\xi - t_{HSN}$.

Our simulations indicate that the increase of $t_P$ with decreasing $I_B$ may not be unique to the SNAP operation. We simulated the operation of a SNSPD after a HSN event for different bias currents (see Ref. [12]) and found that $t_P$ increases from 76 ps at $I_B = I_{SW}$ to 127 ps at $I_B = 0.64 I_{SW}$. The physical process responsible for the abrupt increase in the width and asymmetry of the IRF of SNAPs as $I_B$ approached $I_{AV}$ remains unexplained. The timing jitter may be related to statistical variations occuring within the resistive slab creation time (between $t_{HSN}$ and $t_\xi$), while the increase in the asymmetry with decreasing bias current may be due to the increase in the time required by the current redistributed from the initiating section to suppress the superconducting gap in the secondary sections.

The central result of this paper is the experimental observation that as the bias current of SNAPs was decreased from the device switching current, the device instrument response function (IRF) shifted to longer time delays and became more broad and asymmetric. While we were able to develop a model of the IRF time shift that closely described the experimental data, we could not explain the change in shape of the IRF as the bias current was varied.

The authors thank Andrew Kerman, Hyunil Byun, James Daley, Mark Mondol and Prof. Rajeev Ram for technical support. Detector fabrication and modeling was supported by the Center for Excitonics (Award # DE-SC0001088). Measurements and work at MIT Lincoln Laboratory were sponsored by the United States Air Force under Air Force Contract #FA8721-05-C-0002. Opinions, interpretations, recommendations and conclusions are those of the authors and are not necessarily endorsed by the United States Government. This work was completed while Prof. K. K. Berggren was on sabbatical at Delft University of Technology, and supported by the Netherlands Organization for Scientific Research.

1. M. Ejrnaes, R. Cristiano, O. Quaranta, S. Pagano, A. Gaggero, F. Mattioli, R. Leoni, B. Voronov, and G. Gol'tsman, Applied Physics Letters **91** (26), 262509 (2007).
2. F. Marsili, F. Najafi, E. Dauler, F. Bellei, X. L. Hu, M. Csete, R. J. Molnar, and K. K. Berggren, Nano Lett **11** (5), 2048-2053 (2011).
3. G. N. Gol'tsman, O. Okunev, G. Chulkova, A. Lipatov, A. Semenov, K. Smirnov, B. Voronov, A. Dzardanov, C. Williams, and R. Sobolewski, Applied Physics Letters **79** (6), 705 (2001).
4. J. Zhang, W. Słysz, A. Pearlman, A. Verevkin, R. Sobolewski, O. Okunev, G. Chulkova, and G. Gol'tsman, Physical Review B **67** (13) (2003).
5. J. A. O'Connor, M. G. Tanner, C. M. Natarajan, G. S. Buller, R. J. Warburton, S. Miki, Z. Wang, S. W. Nam, and R. H. Hadfield, Applied Physics Letters **98** (20), 201116 (2011).
6. E. A. Dauler, B. S. Robinson, A. J. Kerman, J. K. W. Yang, K. M. Rosfjord, V. Anant, B. Voronov, G. Gol'tsman, and K. K. Berggren, IEEE Trans. Appl. Supercond. **17** (2), 279-284 (2007).
7. H. Takesue, S. W. Nam, Q. Zhang, R. H. Hadfield, T. Honjo, K. Tamaki, and Y. Yamamoto, Nature Photonics **1** (6), 343-348 (2007).
8. M. J. Stevens, R. H. Hadfield, T. Gerrits, T. S. Clement, R. P. Mirin, and S. W. Nam, Journal of Modern Optics **56** (2-3), 358-363 (2009).
9. As the standard deviation of the number of events in a bin of a histogram is proportional to the square root of the total number of events in the histogram, a large number of photodetection delay samples have to be acquired in order to precisely characterize the timing performance of a detector. The time required to acquire a certain number of time delay samples increases with decreasing detection efficiency of the detector.

[10] F. Marsili, F. Najafi, C. Herder, and K. K. Berggren, Applied Physics Letters **98** (9) (2011).

[11] This offset ensures that the time delay corresponding to the maximum of the IRF is non-negative, while neglecting the effect of the lengths of the electrical and optical paths (mainly RF cables and optical fibers).

[12] See supplementary material at [URL] for: IRF of SNAPs in arm-trigger regime; Discussion of our definition of the MLD and of $t_P$; Bias-dependence of $t_P$ for a SNSPD.

[13] M. Ejrnaes, A. Casaburi, R. Cristiano, O. Quaranta, S. Marchetti, N. Martucciello, S. Pagano, A. Gaggero, F. Mattioli, R. Leoni, P. Cavalier, and J. C. Villégier, Applied Physics Letters **95** (13), 132503 (2009).

[14] We attributed the discrepancy between our results and those of Ref. 13 to the fact that the data reported in Ref. 13 were obtained: (1) with 5-SNAPs, which we also studied without observing avalanche regime operation; (2) with 60-ps-wide laser pulses, resulting in a higher set-up jitter than in our measurements; and (3) by biasing the devices further away from $I_{SW}$ than in our experiment.

[15] A. D. Semenov, G. N. Gol'tsman, and A. A. Korneev, Physica C **351** (4), 349-356 (2001).

[16] A. Semenov, A. Engel, H. W. Hübers, K. Il'in, and M. Siegel, Eur. Phys. J. B **47** (4), 495-501 (2005).

[17] The time difference $t_{HSN} - t_{FPD}$ was set by the propagation times of the signals through the optical and electrical paths of our set up, see Figure 2a.

[18] M. Tinkham, in *Introduction to Superconductivity*, McGraw Hill Inc., New York, 1996, Chap. 11.

# Timing performance of 30-nm-wide superconducting nanowire avalanche photodetectors: Supplementary Material

F. Najafi[1†], F. Marsili[1†], E. Dauler[2], R. J. Molnar[2], K. K. Berggren[1‡]

[1] *Department of Electrical Engineering and Computer Science, Massachusetts Institute of Technology, 77 Massachusetts Avenue, Cambridge, Massachusetts 02139, USA*

[2] *Lincoln Laboratory, Massachusetts Institute of Technology, 244 Wood St., Lexington, Massachusetts 02420, USA*

[†]*these authors contributed equally.*

[‡]*corresponding author:* berggren@mit.edu.

**Table of Contents**

# IRF of SNAPs in arm-trigger regime

When biased below the avalanche current ($I_{AV}$), 3- and 4-SNAPs operated in arm-trigger regime [1]. In this regime the devices did not operate as single-photon detectors because more than one hotspot nucleation (HSN) event was necessary to trigger a detector pulse (2 HSN events for 3-SNAPs and 2 or 3 HSN events, depending on the bias current, for 4-SNAPs). Therefore, as the bias current ($I_B$) was decreased below $I_{AV}$, the devices transitioned from operating as 3-SNAPs (4-SNAPs) biased slightly above $I_{AV}$ to operating as pseudo 2-SNAP (pseudo 3- or 2-SNAPs, depending on the bias current) biased close to the switching current ($I_{SW}$).

Figure SM 1a shows the instrument response function (IRF) of a 3-SNAP for $I_B$ ranging from $I_{SW}$ to $0.52I_{SW}$. The IRF became wider and more asymmetric as $I_B$ was decreased from $I_{SW}$ to ~$0.8I_{SW}$. For $I_B$ slightly below ~$0.8I_{SW}$, the IRF abruptly changed shape and became approximately as narrow and symmetric as the IRF measured for $I_B \sim I_{SW}$. As $I_B$ was decreased further, the IRF became again wider and more asymmetric. Figure SM 1b and c show a quantitative characterization of the shape of the IRF of 3- and 4-SNAPs in terms of its width (jitter) and asymmetry. The abrupt changes in the shape of the IRF as $I_B$ was decreased can be explained with the arm-trigger-regime model, as discussed in Ref. [1].

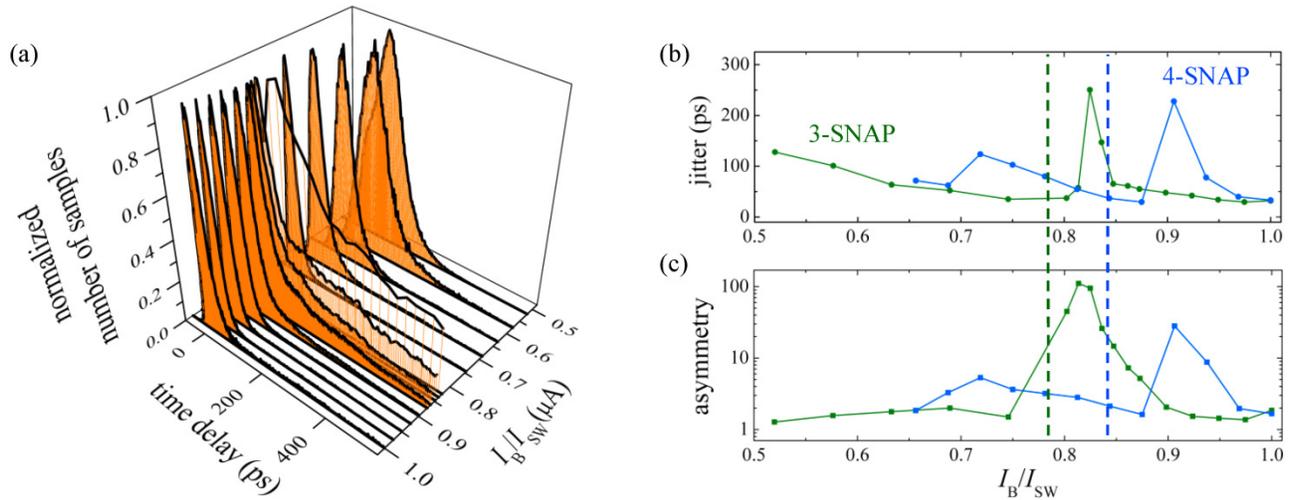

**Figure SM 1. a.** IRF (normalized by the maximum of each trace) of a 30-nm-wide 3-SNAP for $I_B$ ranging from $I_{SW}$ to $0.52I_{SW}$. **b, c.** Jitter (b, defined as the FWHM of the IRF) and IRF asymmetry (c) of a 30-nm-wide 3-SNAP (green, $I_{SW}$ = 17.9 µA) and 4-SNAP (blue, $I_{SW}$ = 25.6 µA). The IRF asymmetry was defined as the ratio between the IRF tails (experimentally defined as the time between 90% and 10% of the IRF maximum) before and after the maximum-likelihood delay.

**Discussion of our definition of the MLD and of $t_P$**

We adopted $t_{SNAP}$ as a reference to measure the MLD to maximize the count rate (and then minimize the acquisition time) and to minimize the counts due to the electrical noise when measuring the IRF ("false counts", see Ref. [1] ).

We used low-noise 3-GHz-bandwidth amplifiers to read out the SNAPs. Therefore the rise time [2] of the measured SNAP photoresponse pulse was limited by the bandwidth of our amplifiers. Figure SM 2 shows the measured averaged voltage pulse of a 2-SNAP at different bias currents. Due to the bandwidth limitation, we observed a bias-independent delay (~305ps; see inset of ) between the times at which the rising edge of the SNAP photoresponse pulse reached 50% and 95 % of the peak value ($t_{SNAP}$ and $t_{95\%}$). This constant offset between $t_{SNAP}$ and $t_{95\%}$ allowed us to measure the current-dependent behavior of $t_{95\%}$ by measuring the current-dependent behavior of $t_{SNAP}$.

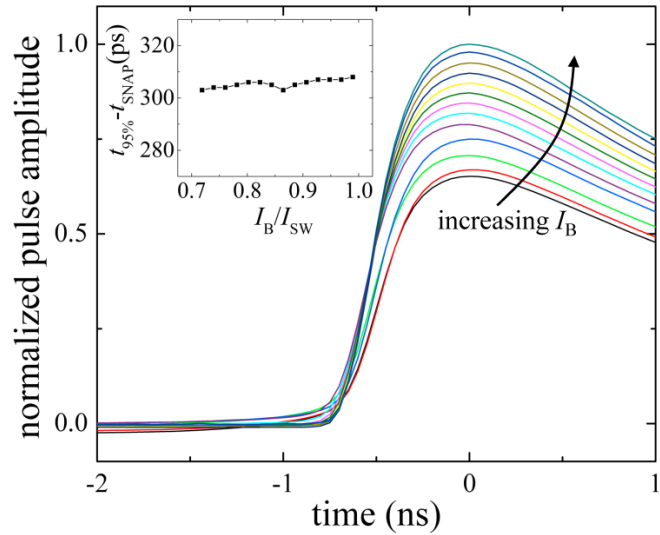

**Figure SM 2.** Measured voltage pulse (averaged over ~ 5000 traces; normalized to the pulse amplitude at $I_B=0.99I_{SW}$) of a 30-nm-wide 2-SNAP for $I_B$ ranging from $0.99I_{SW}$ to $0.72I_{SW}$ by steps of ~$0.02I_{SW}$. The time at which the pulse reached its maximum value was set to 0 s. **inset.** Time delay between the 95%-of-maximum ($t_{95\%}$) and 50%-of-maximum transition (at $t_{SNAP}$) of the rising edge of the voltage pulse as a function of normalized bias current $I_B/I_{SW}$. The maximum variation of $t_{95\%}$-$t_{SNAP}$ for $I_B$ ranging from $0.99I_{SW}$ to $0.72I_{SW}$ was 5ps.

Figure SM 3a shows the bias dependence of the MLD (experiment, black curve) and of the detector peak delay $t_P$ (simulation, in color) extracted from the simulated SNAP pulses shown in Figure 3a by using different thresholds on the SNAP pulse as references: $t_{max}$ (red curve), the instant at which the SNAP photoresponse pulse reaches its maximum; $t_{95\%}$ (orange curve), the instant at which the rising edge of the SNAP photoresponse pulse reaches 95% of the pulse peak value; and $t_{SNAP}$ (green curve), the instant at which the rising edge of the SNAP photoresponse pulse reachesb 50% of the pulse

peak value. Changing the reference threshold on the SNAP pulse did not significantly affect the bias dependence of $t_P$ and then our conclusions.

We chose $t_{95\%}$ over $t_{max}$ because the pulses were flat around their maximum, so choosing $t_{max}$ as a time reference to calculate $t_P$ introduced an uncertainty on the value of $t_P$ of the order of tens of ps.

We chose $t_{95\%}$ over $t_{SNAP}$ because, due to the limited bandwidth of our amplifiers (~ 3GHz), the experimental rise time of the SNAP pulses (~ 300ps, see Figure SM 2) was significantly larger than the rise time of the simulated pulses (~100-150ps) shown in Figure 3a, so we expected the low-passed experimental pulses to be less distorted close to the slopeless maximum of the pulse than at 50% of the maximum. To further support our choice, we numerically filtered the simulated pulses shown in Figure 3a by using a low-pass filter with a 3-dB-cut-off frequency of 3 GHz and extracted $t_{95\%}$ and $t_{SNAP}$ from the filtered pulses. Figure SM 3b shows the bias-dependence of $t_{95\%}$ and $t_{SNAP}$ of the pulses shown in Figure 3a and of the filtered pulses. We added a different offset to each curve so that $t_{95\%}$ and $t_{SNAP}$ would be zero at $I_B = I_{SW}$. While $t_{SNAP}$ of the filtered pulses differed from $t_{SNAP}$ of the pulses shown in Figure 3a by $13 \pm 7$ ps, $t_{95\%}$ of the filtered pulses differed from $t_{95\%}$ of the pulses shown in Figure 3a by $4 \pm 4$ ps, which confirmed that $t_{95\%}$ was a more suitable reference than $t_{SNAP}$ to compare the results of our simulations ($t_P$) to the experimental results (the MLD).

Our choice of $t_{95\%}$ was further motivated by the experimental observation that the bandwidth-limited time difference $t_{95\%} - t_{SNAP}$ did not vary with $I_B$, as shown in Figure SM 2.

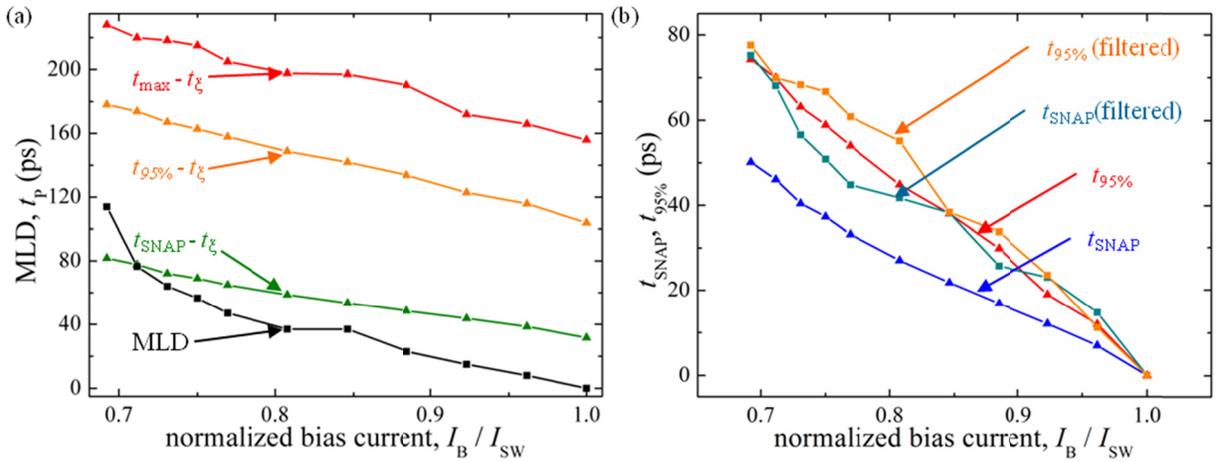

**Figure SM 3. a.** Experimental MLD vs $I_B$ (black squares) and simulated $t_P$ vs $I_B$ (red, orange and green triangles) for the 2-SNAP of Figure 2b. The values of $t_P$ were calculated by using different thresholds on the SNAP pulse as references: $t_{max}$ (red curve); $t_{95\%}$ (orange curve); and $t_{SNAP}$ (green curve). **b.** $t_{95\%}$ and $t_{SNAP}$ of the pulses shown in Figure 3a (triangles) and of the filtered pulses vs $I_B$ (squares).

## Bias-dependence of $t_P$ for a SNSPD

Figure SM 4a and b show the simulated time evolution after a HSN event (occurring at $t = t_\xi = 0$ s) of the current diverted from a SNSPD to the read out ($I_{out}$, see Figure SM 4a) and of the device resistance ($R_{SNSPD}$, see Figure SM 4b). We repeated the simulation at different values of $I_B$. Figure SM 4c shows $t_P = t_{95\%} - t_\xi$ for the SNSPD as a function of $I_B$.

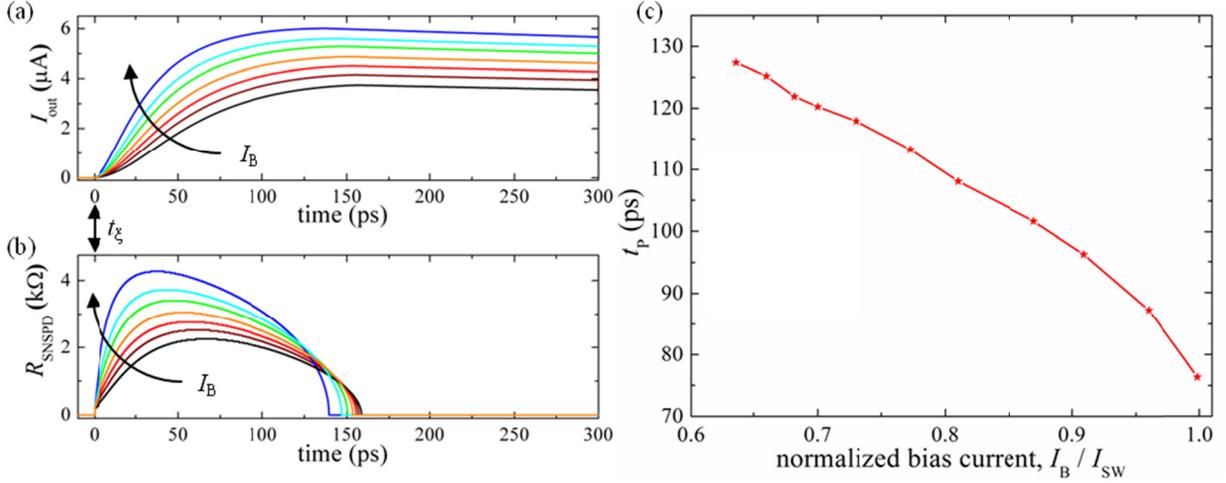

**Figure SM 4. a, b.** Simulated time evolution of $I_{out}$ (a) and $R_{SNSPD}$ (b) after a resistive $\xi$-slab is formed (at time $t_\xi = 0$ s) for an SNSPD biased at $I_B / I_{SW}$= 0.96, 0.91, 0.81, 0.87, 0.81, 0.77, 0.68. The SNSPD had the same kinetic inductance as the 2-SNAP of Figure 3, $L_{SNSPD}$ = 136.5 nH. Black arrows indicate the time at which the resistive $\xi$-slab was formed ($t_\xi$) and the direction of increasing $I_B$. **c.** Simulated $t_P$ vs $I_B$ for the SNSPD of (a) and (b).

## References

[1] F. Marsili, F. Najafi, E. Dauler, X. Hu, M. Csete, R. Molnar, and K. K. Berggren, Nano Lett. **11** (5), 2048 (2011).

[2] F. Marsili, F. Najafi, C. Herder, and K. K. Berggren, Applied Physics Letters **98** (9) (2011).